\documentclass[pre,amsmath,amssymb,twocolumn]{revtex4-1}

\usepackage{graphicx}
\usepackage{dcolumn}
\usepackage{bm}

\usepackage[english]{babel}

\usepackage[utf8]{inputenc}
\usepackage{mathptmx}
\usepackage{etoolbox}
\newcommand{\bi}[1]{Fig.~\ref{fig:#1}}

\usepackage{epsfig}
\usepackage{color}

\usepackage{stackengine}
\usepackage{multirow}

\renewcommand{\(}{\left(} 
\renewcommand{\)}{\right)}

\newcommand{\se}[1]{sec.~\ref{sec:#1}}

\newcommand{\nn}{\nonumber}

\newcommand{\ab}{{\alpha\beta}}
\renewcommand{\P}{\mathcal{P}}
\newcommand{\N}{\mathcal{N}}

\newcommand{\bs}[1]{\boldsymbol{#1}}
\newcommand{\lr}[1]{\langle{#1}\rangle}

\begin{document}

\title{Theory of the asynchronous state of structured rotator networks and its application to recurrent networks of excitatory and inhibitory units}

\author{Jonas Ranft}
\email{jonas.ranft@ens.psl.eu}
\affiliation{Institut de Biologie de l'ENS, Ecole Normale Sup{\'e}rieure, CNRS, Inserm, Universit{\'e} PSL, 46 rue d'Ulm, 75005 Paris, France}
\author{Benjamin Lindner}
\affiliation{Bernstein Center for Computational Neuroscience, Berlin, Philippstra{\ss}e 13, Haus 2, 10115 Berlin, Germany and Physics Department of Humboldt University Berlin, Newtonstra{\ss}e 15, 12489 Berlin, Germany}

\date{\today}

\begin{abstract}
Recurrently coupled oscillators that are sufficiently heterogeneous and/or randomly coupled can show an asynchronous activity in which there are no significant correlations among the units of the network. The asynchronous state can nevertheless exhibit a rich temporal correlation statistics, that is generally difficult to capture theoretically. For randomly coupled rotator networks, it is possible to derive differential equations that determine the autocorrelation functions of the network noise and of the single elements in the network. So far, the theory has been restricted to statistically homogeneous networks, making it difficult to apply this framework to real-world networks, which are structured with respect to the properties of the single units and their connectivity. A particularly striking case are neural networks for which one has to distinguish between excitatory and inhibitory neurons, which drive their target neurons towards or away from firing threshold. To take into account network structures like that, here we extend the theory for rotator networks to the case of multiple populations. Specifically, we derive a system of differential equations that govern the self-consistent autocorrelation functions of the network fluctuations in the respective populations. We then apply this general theory to the special but important case of recurrent networks of excitatory and inhibitory units in the balanced case and compare our theory to numerical simulations. We inspect the effect of the network structure on the noise statistics by comparing our results to the case of an equivalent homogeneous network devoid of internal structure. Our results show that structured connectivity and heterogeneity of the oscillator type can both enhance or reduce the overall strength of the generated network noise and shape its temporal correlations.

\end{abstract}

\maketitle

\section{Introduction}
Groups of interacting oscillators are capable of the prominent phenomenon of \emph{synchronization} -- famous text books have been devoted to the topic \cite{Kur03,PikRos01}. A broad literature identified conditions under which oscillators 
rotate to some degree in synchrony and explored how this is affected by the distribution of eigenfrequencies, by disorder in the coupling coefficients, or by dynamical noise
(see e.g.~\cite{AniAst02,FreSch03,AceBon05,BruHan06,DoiRin06}). However, interacting oscillators can also be in an \emph{asynchronous state} which is far from being a trivial counterpart of the synchronized case -  it is both important in many applications as well as mathematically challenging to describe, and it is what we focus on in this article.  

In neuroscience, networks of interest consist of excitatory and inhibitory  neurons that are sparsely  coupled to each other via chemical synapses \cite{BraSch98,Spo16}. In theoretical studies, networks with sparse random recurrent connectivity display distinct states of activity, among them prominently the asynchronous state, in which the firing of one neuron is not or only very weakly correlated to the firing of other neurons \cite{VanSom96,RenDel10,HelTet14,Ost14}. Interestingly, this type of state is often observed in cortical networks of animals in the attentive state \cite{PouPet08,HarThi11}.  

In the asynchronous state of neural networks, the main statistics of interest is the mean activity and the temporal correlation of spikes within the spike train of a single cell -- as different neurons are not correlated by definition of this state, their cross-correlations can be neglected.
We note that early theories of recurrent networks often focussed predominantly on the mean activity (the firing rate), employing a coarse Poisson (white-noise) approximation for the network-generated fluctuations seen by the single cell \cite{AmiBru97,BruHak99,Bru00}. However, it has been shown that the generated network noise is white only in the limit case of very weak activity \cite{LerSte06,DumWie14} and displays in general pronounced correlations -- in particular if realistic synaptic coupling strengths are taken into account \cite{Ost14,WieBer15,PenVel18}. 

What makes the second problem of the output's autocorrelations mathematically tricky, is their self-consistence \cite{LerSte06,DumWie14}: The correlations in the output of one cell (output correlation) constitute the input-noise correlation for another cell that receives input from this very neuron. One may naively hope that the input correlations will wash out into a temporally uncorrelated (Poissonian) total input when we sum over the many statistically independent input spike trains arriving at the synapses of the considered neuron, but it is relatively easy to show that this is not the case \cite{Lin06}. Hence, when considering spike generation for a neuron embedded in the recurrent network, we do not know the statistics of the network noise that this neuron is subject to; we only know that this input noise should have the same correlation function, or equivalently the same power spectrum as the spike train (the output) generated by the very same neuron. This problem of determining self-consistently the correlation statistics in recurrent networks of neural oscillators has been pursued in three different ways in the past.

First, we can turn the condition of self consistence into an iterative single-neuron simulation scheme \cite{Mar00,LerSte06,LerUrs06,DumWie14}, in which we replace the recurrent input by a surrogate Gaussian noise. In contrast to the diffusion approximation, which employs white Gaussian noise ~\cite{AmiBru97,BruHak99,Bru00}, here temporal correlations of the input noise are taken into account by changing iteratively the power spectrum of the input noise according to the output spectrum of the previous iteration. This can be repeated until input and output spectra match to some degree of accuracy. The simplest version of this procedure works only for a homogeneous network  \cite{LerSte06,DumWie14} but it can be generalized to heterogeneous setups as well \cite{PenVel18}. This approach provides efficient numerical schemes to generate spike trains with power spectra that are in many cases indistinguishable from those measured in large recurrent networks. However, it does not give us a full theory of the self-consistent autocorrelations as the scheme ultimately requires stochastic simulations.

Secondly, it is possible to extend the classical diffusion approximation via the Fokker-Planck approach for recurrent networks of integrate-and-fire neurons. In a more accurate description of the network noise, additional variables via a Markovian embedding account for the self-consistent correlations of the input noise \cite{VelLin19}. The white noise is here replaced by a multidimensional Ornstein-Uhlenbeck process the coefficients of which have to be found from the self-consistency condition for the autocorrelations of the spike train. The resulting high-dimensional Fokker-Planck equation is difficult to treat both analytically and numerically.

A third approach to the problem of the self-consistent correlations of neural oscillators is a radical simplification of the neuron model. Instead of an integrate-and-fire model as used in most of the aforementioned papers, one can e.g.~use rate units as done in the pioneering study by Sompolinsky, Crisanti and Sommer \cite{SomCri88} and derive a single differential equation giving access to the self-consistent autocorrelation function of the units (as comprehensively explained in~\cite{SchGoe16}). This ansatz has more recently  been generalized to take into account heterogeneity in the network \cite{AljSte15,KadSom15,MasOst17}. 

Another approach employing a simplified model has been suggested by van Meegen and Lindner \cite{VanLin18} who studied a randomly connected network of phase rotators and derived equations for the autocorrelation function of the network noise and of the single units.
We have, quite recently, extended their theory to the case of additional (common or individual) dynamic noise \cite{RanLin22}. In both studies, the model was restricted to have a homogeneous random connectivity, a setting in which it is not possible  to strictly define excitatory and inhibitory units. Here we present another extension of the theory, namely, for networks comprising populations of phase rotators with distinct properties. We test the theory for the important special case of recurrent networks of excitatory and inhibitory units, a network of two populations that differ in their statistics of eigenfrequencies and connectivity. We discuss how the heterogeneity of the network affects the temporal correlations of the network noise and    
the correlations of the single units and compare specifically to the case of a homogeneous network.

Our paper is organized as follows. In the next section, we introduce the model for a network of multiple populations of rotators  and develop the general theory for the network noise correlation functions. In \se{EI} we apply this general theory to the case of EI networks in the balanced state and highlight the differences for the correlation statistics (correlation functions and power spectra) to the statistics that would be observed in a homogeneous network with statistically identical but randomized (unstructured) single-unit properties and connectivities. Finally, in \se{discussion} we summarize our results and  discuss open problems for future research.

\section{General theory for multiple populations}
\label{sec:general}

We consider the following dynamical equation for phase variables $\theta^\alpha_m$ of $P$ distinct populations $\alpha\in\P=\{A,B,\ldots\}$, $m\in\N_\alpha=\{1,\ldots,N_\alpha\}$:
\begin{equation}
\dot \theta^\alpha_m = \Omega^\alpha_m + \sum_{\beta\in\P} \sum_{n\in\N_\beta} K^{\ab}_{mn} F_\ab(\theta^\beta_n(t))
\end{equation}
where in all generality also the coupling functions $f_\ab$ may depend both on pre- and postsynaptic populations. This amounts to a total number of $N=\sum_\alpha N_\alpha$ equations for the $N$ neurons in the heterogeneous network. 

We assume that all values of intrinsic frequencies $\Omega^\alpha_m$ and the coupling coefficients $K^{\ab}_{mn}$ are independently drawn for all $N$ units and $N(N-1)$  connections between distinct neurons, respectively, excluding self-connections. The heterogeneity of the network is captured by potentially distinct statistics of the underlying distributions, for each population $\alpha$ in the case of the intrinsic frequencies and for pairs of populations $\ab$ in the case of the coupling coefficients. Note that the block matrices $K^{\ab}_{mn}$ together contain the $N\times N$ recurrent connections among all neurons (self-connections being set to zero). We furthermore take into account potential heterogeneity in the coupling functions $F_\ab$ between (not necessarily distinct) populations $\beta$ and $\alpha$. 

Specifically, we consider the intrinsic frequencies to be Gaussian distributed with respective mean values $\lr{\Omega^\alpha_m}=\Omega^\alpha_0$ and variances $\lr{(\Omega^\alpha_m-\Omega^\alpha_0)^2} = \tilde\sigma_\alpha^2$. For the theory, we do not have to assume a particular distribution for the coupling coefficients $K^\ab_{mn}$ but characterize them exclusively by their first- and second-order statistics:
\begin{subequations}
\begin{align}
\lr{K^\ab_{mn}} &= \kappa^\ab_1 \\
\lr{K^\ab_{mn}K^{\gamma\delta}_{op} } &= \kappa^\ab_1\kappa^{\gamma\delta}_1 
+ \kappa^\ab_2 \delta_{\alpha\gamma}\delta_{\beta\delta}\delta_{mo}\delta_{np} \\
\label{eq:Kvar_individual}
\lr{(K^\ab_{mn})^2 } &= (\kappa^\ab_1)^2 + \kappa^\ab_2
\end{align}
\end{subequations}
Here and in the following, the angular brackets represent averages over different realizations of the network, the intrinsic frequencies, and initial conditions. Specific choices for the coupling matrices are introduced and discussed further below.

In the following, we will assume that the presynaptic phases are uniformly distributed, as can be expected for an asynchronous state. We can then introduce shifted intrinsic frequencies $\omega^\alpha_m$ and coupling functions $f_\ab$ to absorb any finite mean input. With 
\begin{subequations}
\begin{align}
F^\ab(\theta) &=  \sum_l A^\ab_l e^{il\theta} = A^\ab_0 +  f_\ab(\theta) \\
f_\ab(\theta) &= \sum_{l\neq 0} A^\ab_l e^{il\theta} 
\end{align}
\end{subequations}
we obtain
\begin{equation}
\label{eq:rescaled_evol_eq}
\dot \theta^\alpha_m = \omega^\alpha_m + \sum_{\beta\in\P} \sum_{n\in\N_\beta} K^{\ab}_{mn} f_\ab(\theta^\beta_n(t)) \, ,
\end{equation}
where
\begin{subequations}
\label{eq:shifted_freqs}
\begin{align}
\label{eq:new_mean_freq}
\lr{\omega^\alpha_m} &= \Omega^\alpha_0 +  \sum_{\beta\in\P} \sum_{n\in\N_\beta} \kappa^\ab_1 A^\ab_0 \equiv  \omega^\alpha_0 \\
\lr{(\omega^\alpha_m- \omega^\alpha_0)^2} &= \tilde\sigma_\alpha^2 + \sum_{\beta\in\P} \sum_{n\in\N_\beta} \kappa^\ab_2 (A^\ab_0)^2 \equiv \sigma^2_\alpha
\end{align}
\end{subequations}

Of particular importance is the so-called balanced state in which the recurrent input to the single unit does not diverge in the limit of an infinite network~\cite{}. A simple choice for such a scenario is a vanishing mean value. From the above it can be seen that the recurrent input vanishes when
\begin{equation}
\sum_{\beta\in\P} \sum_{n\in\N_\beta} \kappa^\ab_1 A^\ab_0 =  \sum_{\beta\in\P} N_\beta \kappa^\ab_1 A^\ab_0 \equiv 0 \,.
\end{equation}
These $P$ equations impose constraints on the parameters $N_\beta$, $\kappa^\ab_1$, and $A^\ab_0$. A typical choice is to  scale  the $\kappa^\ab_1$ with inverse powers of $N_\beta$; the above constraints allow to have a finite (in fact vanishing) mean value for infinite populations even when the individual terms $N_\beta \kappa^\ab_1 A^\ab_0$ would diverge~\cite{}. 

In order to develop a self-consistent theory for coupled populations of neurons, we  rewrite the recurrent input and the governing equations for the phases as follows:
\begin{subequations}
\begin{align}
\label{eq:general_evol_eq_w_noise}
\dot \theta^\alpha_m &= \omega^\alpha_m + \xi^\alpha_m \\
\label{eq:def_xi_a}
\xi^\alpha_m &= \sum_{\beta\in\P} \xi^\ab_m \ , \quad \xi^\ab_m = \sum_{n\in\N_\beta} K^{\ab}_{mn} f_\ab(\theta^\beta_n(t))
\end{align}
\end{subequations}
In line with previous theories~\cite{SomCri88,VanLin18} (but see~\cite{RanLin22}), we will  now assume that in the asynchronous state the $\xi^\ab_m$ can be approximated as stochastic noise processes with Gaussian statistics and given temporal correlations. Exactly these statistics, captured by the correlation functions of the $\xi^\ab_m$, we want to determine self-consistently, which eventually will allow us to describe the statistics of the network without solving the full (deterministic) $N$ equations of the network. The following derivations, shown for completeness and to make our presentation self-contained, are similar in spirit to what was done in refs.~\cite{VanLin18,RanLin22} and we refer the interested reader to these works for more detailed discussions of the steps involved.

We first note that the auto-correlation function $C^\alpha_\xi(\tau) = \lr{\xi^\alpha_m(t+\tau)\xi^\alpha_m(t)}$ of the combined input $\xi^\alpha_m$ to population $\alpha$ is given by a simple sum over the auto-correlation functions $C^\ab_\xi(\tau)=\lr{\xi^\ab_m(t+\tau)\xi^\ab_m(t)}$ of the individual inputs $\xi^\ab_m$. According to Eq.~\eqref{eq:def_xi_a},
\begin{align}
C^\alpha_\xi(\tau) &=  \sum_{\beta\in\P}\sum_{\gamma\in\P} \lr{\xi^\ab_m(t+\tau)  \xi^{\alpha\gamma}_m(t)} \nn \\
&= \sum_{\beta\in\P}\lr{\xi^\ab_m(t+\tau)\xi^\ab_m(t)} =  \sum_{\beta\in\P}C^\ab_\xi(\tau) \, ;
\end{align}
in the second line, we used additionally that the sum over $\gamma$ reduces to one term because for $\beta\neq\gamma$
\begin{align}
 &\lr{\xi^\ab_m(t+\tau)  \xi^{\alpha\gamma}_m(t)} \nn\\
 &= \sum_{n\in\N_\beta}  \sum_{n'\in\N_\gamma} \lr{K^{\ab}_{mn} K^{\alpha\gamma}_{mn'} f_\ab(\theta^\beta_n(t+\tau))   f_{\alpha\gamma}(\theta^\gamma_{n'}(t)) } \nn \\ 
&= \sum_{n\in\N_\beta}  \sum_{n'\in\N_\gamma} \lr{K^{\ab}_{mn} K^{\alpha\gamma}_{mn'}} \lr{ f_\ab(\theta^\beta_n(t+\tau))} \lr{f_{\alpha\gamma}(\theta^\gamma_{n'}(t)) }\nn \\ 
&=\sum_{n\in\N_\beta}  \sum_{n'\in\N_\gamma} \kappa^\ab_1\kappa^{\gamma\delta}_1  \cdot 0 \cdot 0 = 0 \, ,
\end{align}
where $\lr{ f_\ab(\theta^\beta_n)}=0$ follows from averaging over random, uniformly distributed $\theta^\beta_n$ (note that $f_\ab(\theta) = \sum_{l\neq 0} A^\ab_l e^{il\theta} $). 

Turning to the correlation functions $C^\ab_\xi(\tau)$, we find
\begin{align}
& \lr{\xi^\ab_m(t+\tau)  \xi^\ab_m(t)} \nn\\
 &= \sum_{n\in\N_\beta}  \sum_{n'\in\N_\beta} \lr{K^{\ab}_{mn} K^\ab_{mn'} f_\ab(\theta^\beta_n(t+\tau))   f_\ab(\theta^\beta_{n'}(t)) } \nn \\ 
&= \sum_{n\in\N_\beta}  \sum_{n'\in\N_\beta} \lr{K^{\ab}_{mn} K^\ab_{mn'}} \lr{ f_\ab(\theta^\beta_n(t+\tau))  f_\ab(\theta^\beta_{n'}(t)) } \nn \\ 
&= \sum_{n\in\N_\beta}   \lr{{K^{\ab}_{mn}}^2} \lr{ f_\ab(\theta^\beta_n(t+\tau))  f_\ab(\theta^\beta_{n}(t)) } \nn \\ 
\label{eq:Cab_1}
&\stackrel{\eqref{eq:Kvar_individual}}{=} N_\beta [(\kappa^\ab_1)^2 + \kappa^\ab_2]  \lr{ f_\ab(\theta^\beta_n(t+\tau))  f_\ab(\theta^\beta_n(t)) } \, .
\end{align}
As above, terms with $n'\neq n$ drop out, assuming that $\theta^\beta_n(t_1)$, $\theta^\beta_{n'}(t_2)$ are uncorrelated since we consider an asynchronous state. 

Using the Fourier representation of $f_\ab$ and expressing the phases as integrals of the respective inputs, 
\begin{align}
\theta^\beta_n(t) &= \theta^\beta_n(t_0) + \int_{t_0}^t dt' \dot\theta^\beta_n(t') \nn \\
&= \theta^\beta_n(t_0) + \omega^\beta_n(t-t_0) + \int_{t_0}^t dt' \xi^\beta_n(t') \, ,
\end{align}
one eventually obtains for the remaining average in Eq.~\eqref{eq:Cab_1}
\begin{align}
&\lr{ f_\ab(\theta^\beta_n(t+\tau))  f_\ab(\theta^\beta_n(t)) } \nn\\
&= \sum_{l\neq0}\sum_{l'\neq0} A^\ab_l A^\ab_{l'} \lr{e^{il\theta^\beta_n(t+\tau)} e^{il'\theta^\beta_n(t)} } \nn\\
&= \sum_{l\neq0}\sum_{l'\neq0} A^\ab_l A^\ab_{l'} \underbrace{\lr{e^{i(l+l')\theta^\beta_n(t_0)}}}_{\delta_{l,-l'}}  \lr{e^{il\int_{t_0}^{t+\tau} dt' \dot\theta^\beta_n(t')} e^{il' \int_{t_0}^t dt' \dot\theta^\beta_n(t') } } \nn\\
&= \sum_{l\neq0} |A^\ab_l|^2 \lr{e^{i\omega^\beta_n l\tau}} \lr{e^{il\int_t^{t+\tau} dt' \xi^\beta_n(t')}  } \nn \\
\label{eq:f_avg_d}
&= \sum_{l\neq0} |A^\ab_l|^2 \Phi_\beta(l\tau) e^{-\frac{l^2}{2}\lr{[y^\beta_n(\tau)]^2}}
\end{align}
In the last step, we introduced the characteristic function of the distribution over the (effective) intrinsic frequencies $\omega^\beta_n$ of population $\beta$, 
\begin{equation}
\Phi_\beta(l\tau) = \lr{e^{i\omega^\beta_n l\tau}} \, ,
\end{equation}
and assumed that the integrated recurrent input (or integrated network noise) 
\begin{equation}
y^\beta_n(\tau;t) = \int_t^{t+\tau} dt' \xi^\beta_n(t')
\end{equation} 
is Gaussian-distributed with zero mean, which implies $\lr{e^{ily^\beta_n(\tau)}} =  e^{-\frac{l^2}{2}\lr{[y^\beta_n(\tau)]^2}}$.

Using furthermore 
\begin{align}
&\int_t^{t+\tau} dt' \int_t^{t+\tau} dt'' \lr{\xi^\beta_n(t') \xi^\beta_n(t'')} \nn\\
&= 2 \int_0^\tau dt (\tau-t) \lr{\xi^\beta_n(t+\tau) \xi^\beta_n(t)} \nn\\
&= 2 \int_0^\tau dt (\tau-t) C^\beta_\xi(t) 
\end{align}
and the definition 
\begin{equation}
\Lambda_\alpha(\tau) = \int_0^\tau dt (\tau-t) C^\alpha_\xi(t) 
\end{equation}
we eventually get the following system of ordinary differential equations for the $\Lambda_\alpha(\tau)$:
\begin{align}
\label{eq:key_result}
\ddot \Lambda_\alpha(\tau) =  \sum_{\beta\in\P}  N_\beta [(\kappa^\ab_1)^2 + \kappa^\ab_2] \sum_{l\neq0} |A^\ab_l|^2 \Phi_\beta(l\tau) e^{-l^2\Lambda_\beta(\tau)} \, ,
\end{align}
the solutions of which give the network-noise autocorrelation according to 
\begin{equation}
\label{eq:Cx_alpha_from_Lambda}
C^\alpha_\xi(\tau) = \ddot \Lambda_\alpha(\tau) \, .
\end{equation}

Eq.~\eqref{eq:key_result} is the key result of our paper. We have reduced the $N$ equations for all the neurons in the nework to $P$ coupled second-order ordinary differential equations for the functions $\Lambda_\alpha(\tau)$ that provide the self-consistent network statistics. In the previously considered case of a homogeneous network with $P=1$, the set of equations reduces to a single one that has been derived by van Meegen and Lindner~\cite{VanLin18}. The same reduction applies when all statistics for the different subpopulations are identical, i.e.,  all parameters become independent of $\alpha$ and $\beta$  ($\kappa^\ab_1=\kappa_1$, ...).

Besides the correlation function of the network noise, we can also determine the correlation statistics of the single rotators. Using the phase evolution given by Eq.~\eqref{eq:general_evol_eq_w_noise}, it is relatively straightforward to calculate the correlation function for the pointer $x^\alpha_m(t)=e^{i\theta^\alpha_m(t)}$ for rotator $m$ of population $\alpha$ with effective intrinsic frequency $\omega^\alpha_m$. (Note that the dynamics $\dot\theta = \omega + \xi$, where $\xi$ is a noise, corresponds to the Kubo oscillator~\cite{Gar85}, with the added complication that the noise is determined self-consistently here.) Along the lines of the derivation of $C^\alpha_\xi$, one eventually obtains
\begin{align}
C_{x^\alpha_m}(\tau) = e^{i\omega^\alpha_m \tau - \Lambda_\alpha(\tau)} \ .
\end{align}
Often, it makes sense to pool the data of the rotators of one type. This corresponds to an average over the autocorrelation functions of the pointers $x^\alpha_m(t)$ of the specific subpopulation. In the limit of large $N_\alpha$, this approaches an average over the frequencies $\omega^\alpha_m$,
\begin{align}
C^\alpha_{x}(\tau) &= \lr{ e^{i\omega^\alpha_m \tau - \Lambda_\alpha(\tau)}}_{\bs{\omega^\alpha}} = \Phi_\alpha(\tau)e^{- \Lambda_\alpha(\tau)} \ ,
\end{align}
where $\Phi_\alpha(\tau)$ is the characteristic function of the effective intrinsic frequency distribution of population $\alpha$. We note that this averaged correlation function is also proportional to the autocorrelation function of a subpopulation activity
\begin{equation}
X^\alpha(t) = \sum_{n \in \N_\alpha} x^\alpha_n(t) \, , 
\end{equation}
because for uncorrelated rotators, one finds readily 
\begin{align}
C^\alpha_X(\tau) &= \sum_{n \in \N_\alpha} C_{x^\alpha_n}(\tau) \approx N_\alpha C^\alpha_{x}(\tau) \, .
\end{align}
Here, we used again that for large $N_\alpha$ the sampling over distinct rotators corresponds to an ensemble average over the effective frequencies.

Instead of correlation functions with pronounced oscillatory components, it can be instructive to consider power spectra of the observables. These are here obtained by numerical Fourier transformation of the autocorrelation functions, according to the Wiener-Khinchin theorem~\cite{Gar85}, $S(\omega) = 2  \Re \int_0^\infty d\tau C(\tau) e^{-i\omega\tau}$.

\section{Application to an excitatory-inhibitory network in the balanced state}
\label{sec:EI}

As an example application, we consider an excitatory-inhibitory (E-I) network comprising two populations $\P=\{E,I\}$. Units are connected with probability $p$ and---if connected---possess a fixed synaptic connection strength $j_\ab$ that depends on pre- and postsynaptic neuron identity, where $j_{\alpha E}>0$ and $j_{\alpha I}<0$ to reflect the excitatory and inhibitory character of populations $E$ and $I$, respectively. For unconnected neurons, the coupling coefficients are set to zero. 
We can then give the network statistics introduced above in terms of the $j_\ab$ and $p$:
\begin{subequations}
\begin{align}
\kappa^\ab_1 &= p j_\ab \\
\kappa^\ab_2 &= p(1-p) j_\ab^2 \\
\lr{(K^\ab_{mn})^2 } &= p j_\ab^2
\end{align}
\end{subequations}
The self-consistent equations~\eqref{eq:key_result} for the network-noise statistics thus become
\begin{align}
\label{eq:EI_general}
\ddot \Lambda_\alpha(\tau) =  \sum_{\beta\in\P}  p N_\beta j_\ab^2 \sum_{l\neq0} |A^\ab_l|^2 \Phi_\beta(l\tau) e^{-l^2\Lambda_\beta(\tau)} \, .
\end{align}
We remind the reader that the characteristic functions $\Phi_\beta$ reflect the effective distributions of the intrinsic frequencies taking into account the mean recurrent input, see the discussion surrounding Eqs.~\eqref{eq:shifted_freqs}. In particular, we see that the sign of the synaptic coupling does not influence the dynamics directly as only $j_\ab^2$ enters Eq.~\eqref{eq:EI_general}.  To distinguish excitation and inhibition in the framework of our model, we thus need coupling functions $f_\ab$ with non-vanishing Fourier components $A^\ab_0$, giving rise to non-vanishing mean synaptic inputs. For the effective intrinsic frequencies, we consider in the following without loss of generality that $A^\ab_0=1$.

\subsection{Simplification of the two-population theory resulting from the balance condition}

As can be seen from the products $p N_\beta j_\ab^2$ appearing in each term of Eq.~\eqref{eq:EI_general}, the network noise can be expected to be finite and non-zero in the limit of large $N$ if the synaptic strengths $j_\ab$ scale with the average number of connections according to
\begin{equation}
j_\ab = \frac{J_\ab}{\sqrt{pN_\beta}} \, .
\end{equation}
We then have
\begin{align}
\label{eq:EI_mean_freq}
\omega^\alpha_0 &= \Omega^\alpha_0 +  \sqrt{p N_E} J_{\alpha E} + \sqrt{p N_I} J_{\alpha I}  \, 
\end{align}
for the mean effective intrinsic frequency of population $\alpha$. In order to ensure that the intrinsic frequencies remain finite in the limit of large $N$ and constant $p$, we choose the case of balanced input,~i.e.,
\begin{equation}
\label{eq:EI_balance_condition}
J_{\alpha I} = - J_{\alpha E} \sqrt{N_E/N_I} \, .
\end{equation}
The spread of the intrinsic frequencies due to the random connectivity remains finite and is in this case given by
\begin{align}
\label{eq:EI-freq-spread}
\sigma^2_\alpha &= \tilde\sigma_\alpha^2 +  (1-p)  J_{\alpha E}^2 (1 + N_E/N_I) \, .
\end{align}
Note that the effective spread of the frequencies depends on our choice of the connectivity, here of the Erdos-R{\'e}nyi type. If for instance we chose a balanced networks with a fixed in-degree, the intrinsic frequencies would not be affected by the (otherwise random) recurrent connections, in contrast to Eq.~\eqref{eq:EI-freq-spread}.

Let us furthermore assume that the coupling functions only depend on the presynaptic identity,~i.e.,
\begin{equation}
f_\ab(\theta) \equiv f_\beta(\theta) \ \text{or } A^{\alpha E}_l \equiv A^E_l \ ,\ A^{\alpha I}_l \equiv A^I_l 
\end{equation}
For the balanced network, we can then further simplify the equations for $\Lambda_\alpha$ with the substitutions
\begin{equation}
\Lambda_\alpha(\tau) = J_{\alpha E}^2 \lambda(\tau) \, ,
\end{equation}
giving rise to a \emph{single} ordinary differential equation for $\lambda(\tau)$:
\begin{align}
 &\ddot \lambda(\tau) \equiv \frac{\ddot \Lambda_\alpha(\tau)}{J_{\alpha E}^2} =  \sum_{\beta\in\P}   \frac{J_\ab^2}{J_{\alpha E}^2} \sum_{l\neq0} |A^\ab_l|^2 \Phi_\beta(l\tau) e^{-l^2 \Lambda_\beta(\tau)} \nn \\
&=  \sum_{l\neq0} |A^{E}_l|^2 \Phi_E(l\tau) e^{-l^2 J_{EE}^2 \lambda(\tau)} +  \frac{N_E}{N_I}  \sum_{l\neq0} |A^{I}_l|^2  \Phi_I(l\tau) e^{-l^2 J_{IE}^2 \lambda(\tau)} \, .
\end{align}
Remarkably, this means that for a balanced network, the theory for a two-population system is almost as simple as the theory for a single homogeneous population. Notably, the correlation functions $C^E_\xi(\tau)$ and $C^I_\xi(\tau)$ of the network input to populations $E$ and $I$, respectively, are simply scaled versions of each other.

\subsection{Comparison to an unstructured network and the corresponding single-population theory}

The theory for the balanced E-I network derived above takes into account the particular structure of the network, which is comprised of two distinct populations with specific input noise statistics. We can now ask how well such a network would be described by the original theory of van Meegen and Lindner developed for networks devoid of internal structure, i.e.~for a homogeneous network. This begs the question of how such a comparison can be meaningfully conceived. We consider here for comparison a randomized version of the original, structured network where all 
connection strengths $K_{mn}$, with $m,n=1,\ldots,N$, are drawn from the combined and properly weighted distributions of the 
$K^\ab_{mn}$, 
with $m=1,...,N_\alpha$, $n=1,...,N_\beta$ and $\alpha,\beta\in\mathcal{P}$. Since in the theory for homogeneous networks the coupling functions between units were assumed to be the same for all connections, we will restrict the comparison further to the case where $f_E(\theta)=f_I(\theta) \equiv f(\theta)$ for all $\theta$.

To apply the single-population theory and assess the effect of network structure by comparison, we thus consider the effective ``single-population'' network 
\begin{subequations}
\begin{align}
\dot \theta_m &= \omega_m + \xi_m \\
\xi_m &= \sum_n K_{mn} f(\theta_m) 
\end{align}
\end{subequations}
where $m$, $n$ are now indices that run over all $N=N_E+N_I$ units. We consider the effective distribution of the intrinsic frequencies of the homogeneous network to be given by $p_{\rm eff}(\omega) = \frac{N_E}{N}\mathcal{N}(\omega^E_0,\sigma_E) + \frac{N_I}{N}\mathcal{N}(\omega^I_0,\sigma_I)$, that is, by the weighted combination of the frequency distributions of the original populations $E$ and $I$, where we take into account the additional broadening due to the recurrent input expected for each population, see Eq.~\eqref{eq:EI-freq-spread}. Note that one could also consider instead the broadening that would be expected from the unstructured recurrent input $\sum_n K_{mn}F(\theta_n)$ on rotators with intrinsic frequencies $\Omega_m$; however, we choose here to restrict our comparison to the effects related to the self-generated network noise. 

For an unstructured network, the network-noise correlation function $C_\xi(\tau)=\lr{\xi_m(t+\tau)\xi_m(t)}$ is given by 
\begin{subequations}
\begin{align}
\label{eq:EI-1pop-Cxi-Lambda}
C_\xi(\tau) &= \ddot \Lambda(\tau) \\
\label{eq:EI-1pop-Lambda-ODE}
\ddot \Lambda(\tau)    &= K^2 \sum_{l\neq0}  |A_l|^2  \Phi(l\tau)   e^{-l^2  \Lambda(\tau)}
\end{align}
\end{subequations}
according to van Meegen and Lindner~\cite{VanLin18}. Here, $K^2$ is related to the variance of the connectivity matrix $K_{mn}$ that contains the recurrent synaptic weights of all $N$ units, and $\Phi(t)$ is the characteristic function of the distribution of effective intrinsic frequencies 
for both populations combined. 
With the synaptic weights between populations given above, one finds for the variance of the connectivity matrix 
\begin{align}
\lr{K_{mn}^2} &= \frac{1}{N}\(J_{EE}^2 \frac{N_E}{N_I} + J_{IE}^2\) \ .
\end{align}
Our approximation of a two-population network would thus become
\begin{align}
\label{eq:EI-1pop-Lambda-ODE}
\ddot \Lambda(\tau)    &= K^2 \sum_{l\neq0}  |A_l|^2 \(\frac{N_E}{N}\Phi_E(lt) + \frac{N_I}{N}\Phi_I(lt) \)   e^{-l^2  \Lambda(\tau)} \, ,
\end{align}
with $K^2=J_{EE}^2 {N_E}/{N_I} + J_{IE}^2$.

\begin{figure*}
\includegraphics[width=0.8\textwidth]{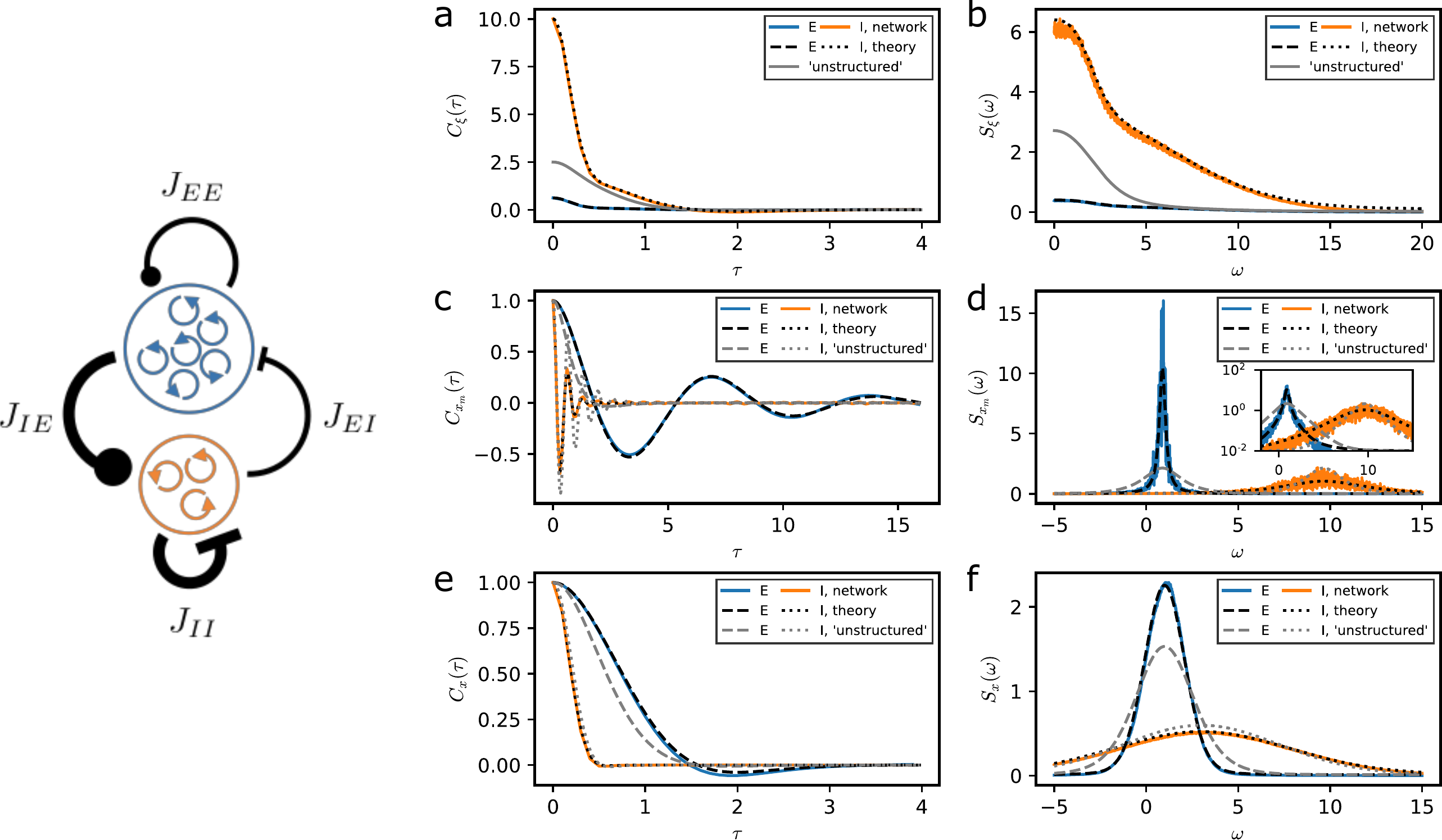}
\caption{\textbf{Network statistics for a balanced E-I network with strong input to the inhibitory population.} Left: sketch of the network with excitatory and inhibitory populations and corresponding connection strengths. Right: measured correlation functions (a,c,e) and power spectra (b,d,f) for network noise (a,b), single rotators (c,d) and population-averaged rotators (e,f). Network parameters: $N_E=800$, $N_I=200$, $p=0.2$, $J_{EE}=0.5$, $J_{IE}=2$, $\Omega^E_0=1$, $\Omega^I_0=3$, $\tilde\sigma_E=\tilde\sigma_I = 0$; simulation parameters: time step $dt=0.01$, simulation length $T=n_0 T_0$ with time window $T_0=1000$ and $n_0=10$, repeated for $n_r=12$ network realizations with newly drawn connectivity matrices; numerical integration parameters: time step $dt=0.01$, $T=1000$. 
\label{fig:1}}
\end{figure*}

From the solution of Eq.~\eqref{eq:EI-1pop-Lambda-ODE}, one obtains the correlation function of the pointer $x_m(t) = e^{i\theta_m(t)}$ of rotator $m$ with effective intrinsic frequency $\omega_m$ according to
\begin{equation}
C_{x_m}(\tau) =  e^{i\omega_m\tau - \Lambda(\tau)} \ .
\end{equation}
(Note that the network-noise autocorrelation function $C_\xi(t)$ is directly given by Eq.~\eqref{eq:EI-1pop-Cxi-Lambda}.) In a strict sense, the theory for an unstructured network does not permit to consider correlation functions at the level of distinct populations. To compare the population-averaged rotator autocorrelation functions predicted by the two-population theory to the `unstructured' network, we can, however, average over the effective intrinsic frequency distributions of populations $E$ and $I$, respectively, such that
\begin{equation}
\label{eq:EI_unstructured_Cx}
C^{E,I}_{x}(\tau) =  \Phi_{E,I}(\tau) e^{ - \Lambda(\tau)} \ .
\end{equation}
Thus comparing the two-population theory to the equivalent `unstructured' network allows one to assess the importance of having a correct description of the self-generated network noise at the level of the individual populations, whereas the first-order effect due to the spread of intrinsic frequencies is taken into account also in the one-population theory.


Let us finally consider the specific case where the connection strengths $J_\ab$ do not depend on the postsynaptic identity, i.e., $J_{EE} = J_{IE}\equiv J$ (and consequently $J_{EI}=J_{II}=-J\sqrt{N_E/N_I}$). For the structured network, with $\Lambda(\tau) = J^2 \lambda(\tau)$, one has
\begin{equation}
C^E_\xi(\tau) = C^I_\xi(\tau)  = \ddot \Lambda(\tau)
\end{equation}
and with $K^2=J^2 N/N_I$ the two-population theory reduces to 
\begin{align}
\label{eq:EI_2pop_same_weight_Lambda}
\ddot \Lambda(\tau)    &= K^2 \sum_{l\neq0}  |A_l|^2 \( \frac{N_I}{N}  \Phi_E(l\tau) +  \frac{N_E}{N}  \Phi_I(l\tau) \) e^{-l^2  \Lambda(\tau)}
\end{align}
which indeed is very similar to the single-population theory Eq.~\eqref{eq:EI-1pop-Lambda-ODE} but interestingly does not exactly correspond to that result. This difference can be understood as follows. Whereas in the unstructured network (Eq.~\eqref{eq:EI-1pop-Lambda-ODE}), the characteristic functions of each population are weighted with the population size, they contribute in the actual two-population network (Eq.~\eqref{eq:EI_2pop_same_weight_Lambda})  with a weight that scales with the synaptic strength. For a typical value of $g\equiv N_E/N_I=4$, the balance condition implies that albeit the inhibitory population is smaller in size, its characteristic function influences  the dynamics of the self-generated noise more strongly.


\subsection{Comparison to network simulations}
Even though we consider the balanced case of an E-I network, we still have some freedom in choosing the synaptic coupling strength for the distinct connections. In the following,  we separately study three such choices that differ in the overall scaling of the synaptic inputs to the inhibitory population compared to the inputs to the excitatory population.\\

\subsubsection{Strong input to the inhibitory population}
\vskip-1em
We start with the case illustrated in \bi{1} on the left: both excitatory and inhibitory inputs to the inhibitory population are chosen to be very strong compared to the inputs to the excitatory population, i.e. $J_{IE} > J_{EE}$ and   $|J_{II}| > |J_{EI}|$, where the latter follows directly from the first with the balance condition~\eqref{eq:EI_balance_condition}. First of all, we remark that all measured correlation functions and power spectra (blue and orange for E and I populations, respectively) are well described by our two-population theory (dashed and dotted black lines), see \bi{1}a-f.

As expected for our choice of input amplitudes, the network noise for neurons in the I population (orange) is stronger than for neurons in the E population (blue), indicated by a larger network-noise correlation function (\bi{1}a) and power spectrum (\bi{1}b); in fact, according to our theory and as confirmed by our simulations, the correlation function of the network noise for the I units is just an upscaled version of that for the E units. In particular, this implies that $C^E_\xi(\tau)$  and  $C^I_\xi(\tau)$ share the same time dependence. We can also compare the network noise correlation functions to the prediction of the single-population theory (grey line), which does \emph{not} share this time course. We note that in the considered case the single-population theory provides correlation functions and power spectra that are in between the corresponding statistics for the E and I populations. This seems to be plausible but has not necessarily to be the  case (see below).

Turning to single-rotator statistics, we show the correlation functions (\bi{1}c) and power spectra (\bi{1}d) for specific rotators, one from each population (blue and orange lines for E and I unit, respectively), and use their eigenfrequencies $\omega_i$ to calculate the corresponding theoretical predictions (dashed and dotted black lines, respectively). While these statistics clearly indicate the presence of intrinsic oscillations, a pronounced damping of the correlation function (\bi{1}c) and, equivalently, a considerable width of the spectral peak (\bi{1}d) is apparent. The damping of the correlation function is related to the phase diffusion of the stochastic oscillator, which in turn is entirely driven by the network noise (see e.g.~Eq.~\eqref{eq:general_evol_eq_w_noise}). Consistent with our choice of strong synaptic couplings onto the I population and, in consequence, a stronger network noise for the units in that population, this damping is considerably stronger for those units. Of note, this difference in damping cannot be captured by the equivalent one-population theory (dashed and dotted grey lines for E and I unit, respectively), as only a single, ``aggregate'' network noise is described by this theory, see also \bi{1}a,b.

We finally show the population-averaged correlation functions (\bi{1}e) and power spectra (\bi{1}f) of the rotators. The intrinsic eigenfrequencies of the rotators differ significantly even within populations---according to Eq.~\eqref{eq:EI-freq-spread}, for excitatory units the standard deviation $\sigma_E=1$ is as large as mean frequency $\Omega^E_0=1$; for inhibitory units, the spread is even more larger ($\sigma_I=4$,  $\Omega^I_0=3$). For this reason, the intrinsic oscillations are much less apparent in the correlation functions, see \bi{1}e. The power spectra are peaked at the respective mean frequencies of both populations, see \bi{1}f, where the spectral width is now a combined effect of the spread of the intrinsic frequencies and the network noise. For the population-averaged rotator statistics, the single-population theory for the equivalent homogeneous network (dashed and dotted grey lines) is closer to the network simulation and two-population theory than for the single rotators (cf.~\bi{1}c,d). Notably the spectral widths of the peaks in the power spectra given by the one-population theory are not identical for both populations (\bi{1}f). We remind the reader that we incorporate an explicit average over the correct intrinsic frequency distribution of each population, see Eq.~\eqref{eq:EI_unstructured_Cx}.  Thus, the improved correspondance of single-population theory for this specific statistics indicates that these spectral widths are to a significant degree determined by the spread of the effective eigenfrequencies.

\subsubsection{Weak input to the inhibitory population}

\begin{figure*}
\includegraphics[width=0.8\textwidth]{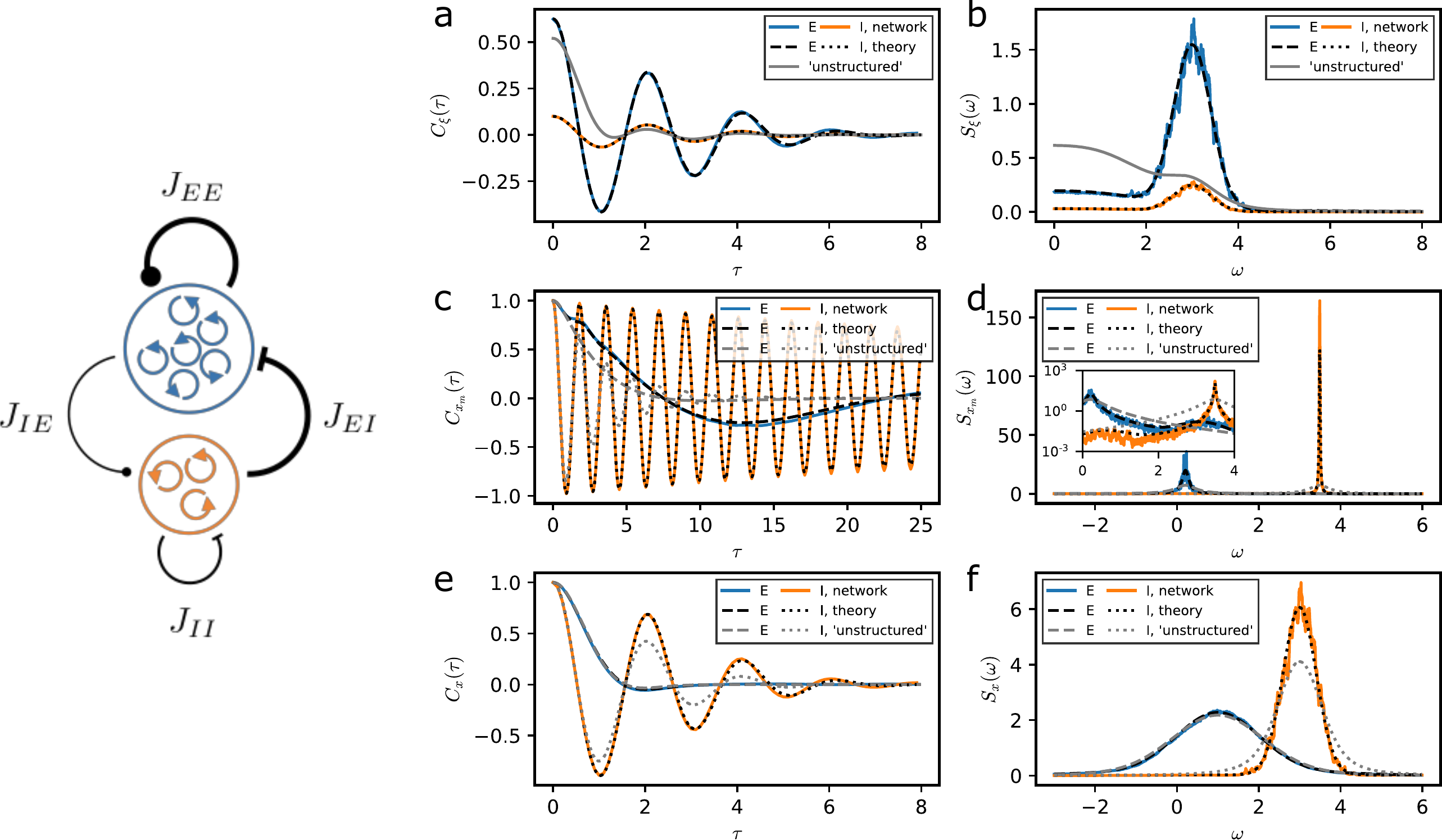}
\caption{\textbf{Network statistics for a balanced E-I network with weak input to the inhibitory population.} Left: sketch of the network with excitatory and inhibitory populations and corresponding connection strengths. Right: measured correlation functions (a,c,e) and power spectra (b,d,f) for network noise (a,b), single rotators (c,d) and population-averaged rotators (e,f). Parameters as in \bi{1} except $J_{IE}=0.2$.
\label{fig:2}}
\end{figure*}

We now discuss the opposite case, where the inputs to the I population are chosen to be smaller than the inputs to the E population, see sketch of \bi{2}, left. Again, the two-population theory (dashed and dotted black lines) nicely captures all measured statistics of the network simulations (blue and orange lines). Since we present the same statistics (correlation functions and power spectra of network simulations, the two-population theory, and the single-population theory) as in the previous case, we will focus in the following on those observations that highlight the difference of the present case from the former. 

First of all, the network noise is now clearly stronger for the excitatory population than for the inhibitory population, see \bi{2}a,b, consistent with a reversal of  `strong' and `weak' synaptic input compared to the situation in \bi{1}. More interestingly, the network noise predicted by the one-population theory exhibits a strong low-frequency component and its statistics does not appear to be a simple interpolation between the statistics of the network noise of excitatory and inhibitory populations. A discrepancy between the one-population theory and the network simulations is again found in the statistics for single rotators (\bi{2}c,d), where the distinct spectral widths of both populations cannot be captured by the approximation of the true network by an unstructured network. We note that the two-population theory describes very well the observed spectra including the weak side peaks, see inset of \bi{2}d. In line with the weak network noise for the I population, the intrinsic oscillation of the I unit is very weakly damped as indicated by the small spectral width (\bi{2}d) and the persistent oscillations in the correlation function (\bi{2}c). For the population averages of the single rotator statistics, the one-population theory again fares better than for the statistics of an individual rotator. However, in the present case of weak input to the inhibitory population, the spread of the intrinsic frequencies of the I population $\sigma_I=0.4$ is considerably smaller than the mean frequency $\Omega^I_0=3$, and the effect of the network noise is more important than in the previous case. This explains the stronger deviation of the population-averaged correlation function and power spectrum predicted by the one-population theory for the I population.

\subsubsection{Inputs of similar strength for E and I units}

\begin{figure*}
\includegraphics[width=0.8\textwidth]{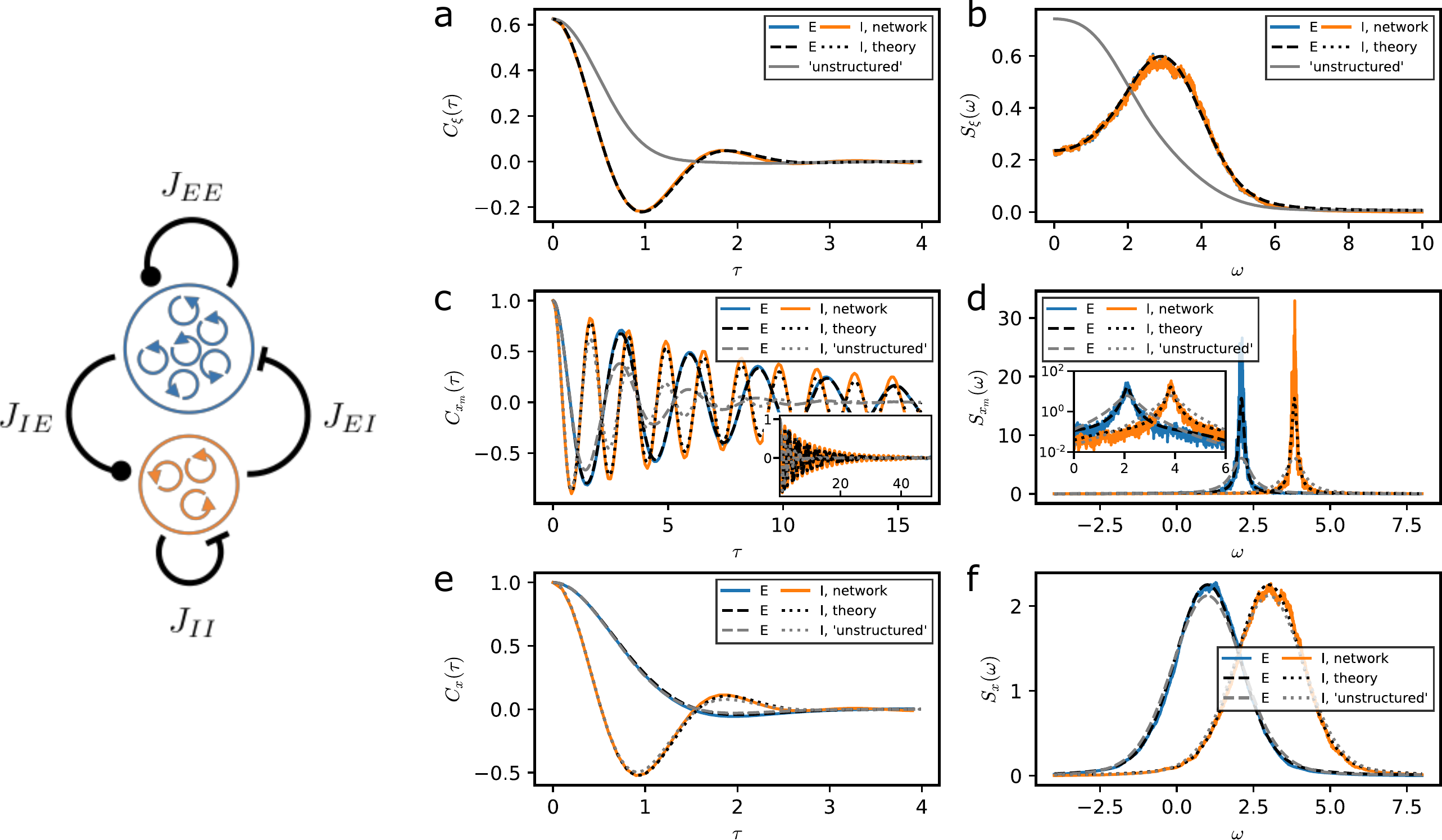}
\caption{\textbf{Network statistics for a balanced E-I network with equal input to the excitatory and the inhibitory populations.} Left: sketch of the network with excitatory and inhibitory populations and corresponding connection strengths. Right: measured correlation functions (a,c,e) and power spectra (b,d,f) for network noise (a,b), single rotators (c,d) and population-averaged rotators (e,f). Parameters as in \bi{1} except $J_{IE}=0.5$.
\label{fig:3}}
\end{figure*}

As a last example, we consider the case where the synaptic input strengths are the same for all units, i.e.~$J_{EE}=J_{IE}$ and $J_{EI}=J_{II}$, see the sketch of \bi{3}, left. As predicted by our two-population theory and confirmed by simulations,  the network noise statistics are then identical for E and I populations (\bi{3}a,b). 

In light of the identical network noise, one might be tempted to think that for the considered situation the theory for the equivalent `unstructured' network should give the same results as the full theory. It is thus instructive to compare the remaining differences between the full heterogeneous network and the one-population theory. As can be seen from \bi{3} (grey lines), the one-population theory fails to describe the observed network noise and rotator statistics correctly: (i) In the network noise spectrum (\bi{3}b), we observe increased power at low frequencies and the absence of a spectral peak at a non-vanishing frequency, in marked contrast to the two-population theory and the network simulations; (ii) in the single-rotator correlation functions and power spectra (\bi{3}c,d), the one-population theory predicts a significantly stronger damping of the correlations, an effect which is also reflected in the population-averaged correlation functions and spectra (\bi{3}e,f).

As already briefly mentioned (cf.~discussion following Eq.~\eqref{eq:EI_2pop_same_weight_Lambda}), the remaining discrepancy between the `structured' and the `unstructured' network is due to the difference in connection weights $J_{\alpha E}$ and $J_{\alpha I}$ of excitatory and inhibitory units, respectively, combined with the difference in intrinsic frequencies. The remaining structure of the considered network is that the faster oscillators (I units) still have much stronger synaptic weight than the slower oscillators (E units)---an effect that cannot be taken into account in the one-population theory. 

\begin{figure*}
\includegraphics[width=0.8\textwidth]{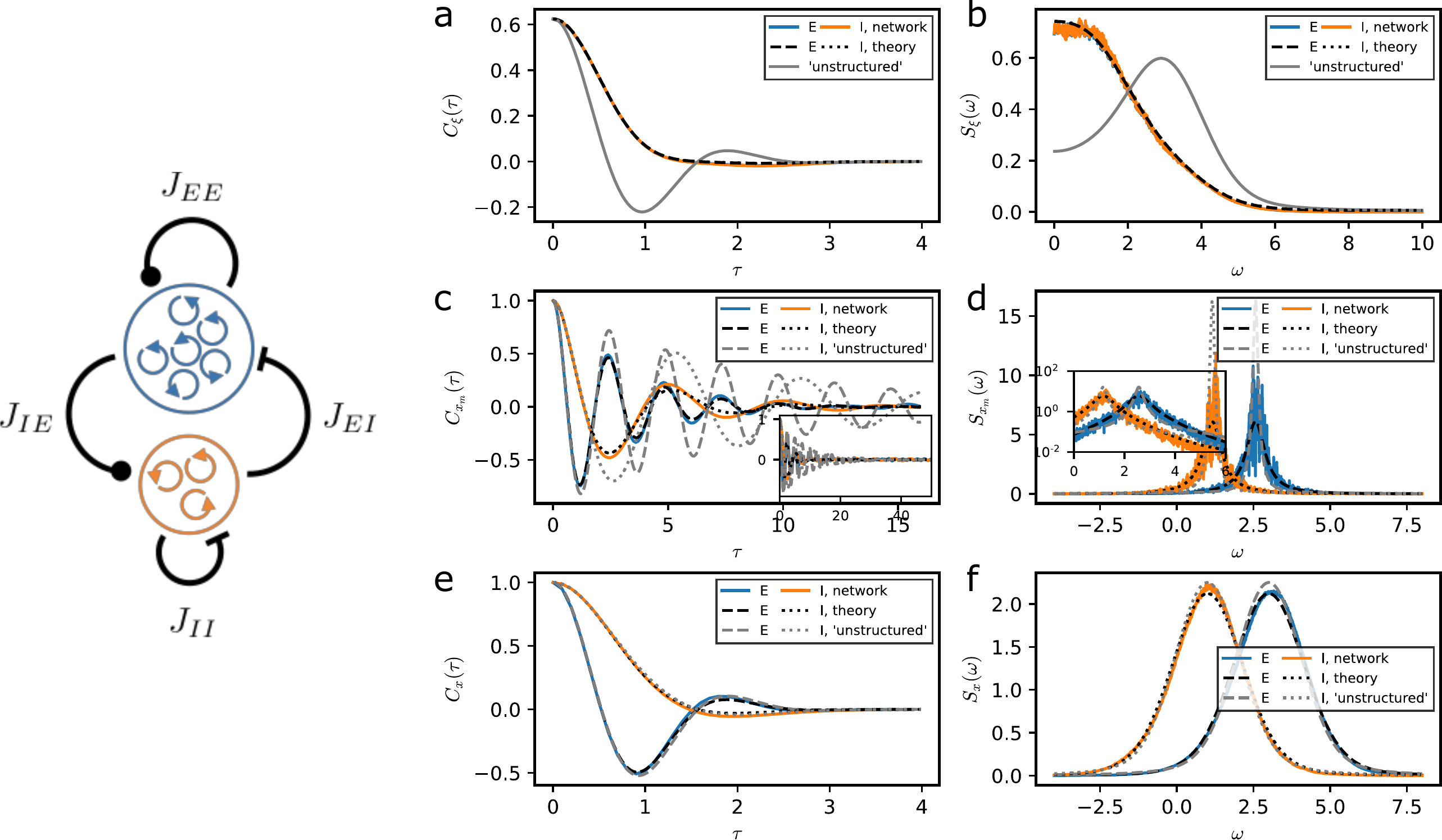}
\caption{\textbf{Network statistics for a balanced E-I network with equal input to the excitatory and the inhibitory populations, with inverted intrinsic frequencies of excitatory and inhibitory units.} Left: sketch of the network with excitatory and inhibitory populations and corresponding connection strengths. Right: measured correlation functions (a,c,e) and power spectra (b,d,f) for network noise (a,b), single rotators (c,d) and population-averaged rotators (e,f). Parameters as in \bi{1} except $J_{IE}=0.5$, , $\Omega^E_0=3$, and $\Omega^I_0=1$.
\label{fig:4}}
\end{figure*}

We further remark that in the case of identical couplings for inhibitory and excitatory units, the contribution of recurrent inputs to the spread of the intrinsic frequencies is the same for both populations (Eq.~\eqref{eq:EI-freq-spread}). For our parameters, the total frequency spread is thus identical ($\sigma_E=\sigma_I=1$) as can also be observed in the single-rotator statistics (\bi{3}c-f). We finally note that while in the present case the one-population theory overestimate the phase diffusion of rotators, this is not a general feature of our setup when coupling strengths $J_{EE}=J_{IE}$ are identical but other parameters can differ. If we, for instance, switch the roles of E and I units with respect to their intrinsic frequencies, see \bi{4}, the predictions for correlation functions and power spectra of the two-population theory become exactly the ones of the one-population theory of \bi{3} and vice versa. In particular, the one-population theory now \emph{underestimates} the broadening of the single-rotator power spectra (\bi{4}d,f).

\section{Summary \& outlook}
\label{sec:discussion}

In this paper we have developed a general theory for the correlation functions of the self-generated network noise and single units in a heterogeneous network of rotators. The heterogeneity was introduced by means of populations with distinct properties of the single elements and distinct connectivities. Our theory, which builds on the work by van Meegen and Lindner, culminated in the system of equations \eqref{eq:key_result}, meaning that in the end we have reduced the problem of calculating the correlations from simulating $N=\sum_{\alpha=1}^P N_\alpha$ elements to solving $P$ ordinary differential equations.

Our results demonstrate, first of all, that the theory works. Despite being strictly valid only in the limit of $N_\alpha\to\infty$ for all $\alpha$ (due to the assumption of Gaussianity for network fluctuations), it works already for comparatively small numbers of units; in our test we used only $N_I=200$ inhibitory units. This means that it can be applied for realistic numbers of oscillators and is useful to capture finite-size effects of the network noise even in cases in which a balance condition is not obeyed.   
 
The theory introduced here could be further explored and generalized in various directions.
One could try to solve the resulting differential equations analytically (for examples in a homogeneous situation, see \cite{VanLin18}), using simplifying assumptions for the coupling statistics and frequency distributions; explicit formulas for the correlation functions and spectra may provide further insights into the effects of network heterogeneity on the generated network noise. 

Beyond the case of two distinct populations, one should explore networks with three or more populations. In particular the diversity of inhibitory cells in the cortex (see e.g.~\cite{JiaShe15}) suggests to study networks with one population of excitatory (pyramidal) cells but two or more types of interneurons (see e.g.~\cite{BerDor21} for a recent computational study of such a network). 


Another extension is to make the rotator coupling more similar to the pulse coupling in recurrent networks of integrate-and-fire models. For homogeneous networks such an approach has been suggested already in  \cite{VanLin18} and it was shown that the self-consistent power spectra of single spike trains could be well approximated if the single integrate-and-fire neurons of the recurrent network operated in a strongly mean-driven regime. We cannot think of a reason why the same approach should not work for a heterogeneous network.  

Furthermore, the inclusion of intrinsic noise sources as in \cite{RanLin22} and of external signals is certainly worth the exploration but may require also complicated corrections to the Gaussian theory put forward here. Inclusion of common stimuli would be interesting also from an information-theoretic perspective. Besides the autocorrelation functions of the autonomous network activity, then the correlation between information-carrying stimuli and the network activity may be calculated permitting to quantify the information transmission by the network.  
All of these are exciting topics for future studies.

\bibliographystyle{apsrev}       
\bibliography{ALL_22-08-24}   

\end{document}